\documentclass[iop]{emulateapj}

\newcommand{\mvc}{v_{\rm c}}

\newcommand{\mdv}{\Delta v}
\newcommand{\dv}{$\Delta v$}

\def\ltp{\left ( \,}

\def\rtp{\, \right  ) }

\newcommand{\nhi}{$N_{\rm HI}$}
\newcommand{\mnhi}{N_{\rm HI}}

\newcommand{\cm}[1]{\, {\rm cm^{#1}}}
\newcommand{\mkms}{{\rm \; km\;s^{-1}}}

\newcommand{\lya}{Ly$\alpha$}

\begin{document}

\title{On the Analysis of DLA Kinematics} 

\author{J. Xavier Prochaska\altaffilmark{1} and
	Arthur M. Wolfe\altaffilmark{2}
}

\altaffiltext{1}{Department of Astronomy and Astrophysics, 
UCO/Lick Observatory;
University of California, 1156 High Street, Santa Cruz, CA  95064;
xavier@ucolick.org}
\altaffiltext{2}{Department of Physics, and 
Center for Astrophysics and Space Sciences, 
University of California, San
Diego, 
Gilman Dr., La Jolla; CA 92093-0424; awolfe@ucsd.edu}

\begin{abstract}
We discuss two mistreatments of damped \lya\ (DLA) kinematic analysis that were first
performed by Haehnelt, Steinmetz, \& Rauch (1998; hereafter HSR98) and have recently
been repeated by Hong et al.\ (2010; astro-ph/1008.4242\_v1,v2;
hereafter H10).  Each
mistreatment led to the improper excising of simulated absorption
profiles.  Specifically, their analyses
are strictly biased against DLA sightlines that have low HI
column density $\mnhi \lesssim 10^{20.5} \cm{-2}$, very
high \nhi\ values, and (for all \nhi) sightlines with low 
\dv\ ($<30 \mkms$ for HSR98; $<20-30\mkms$ for H10).  None of these
biases exist in the observational analysis.  
We suspect these mistreatments compromise the results that followed.  Hopefully
this posting will prevent their repetition in the future.  
\end{abstract}


\section{Introduction}

In 1997, we first measured the kinematic characteristics of a small
sample ($N=17$) of damped \lya\ systems (DLAs) as a means to assess
the dynamical motions of high $z$ galaxies \citep[][hereafter PW97]{pw97}.
We compared these data against simplistic models of protogalaxies
(e.g.\ rotating disks, infalling gas clumps) and favored a thick ($h
\gtrsim 1$\,kpc) disk scenario with rotation speeds $v_{\rm c} \approx 200
\mkms$.  Furthermore, we ruled out standard CDM scenarios that
considered DLAs as rotating disks \citep[e.g.][]{kff96}. These
conclusions hold today.

The following year, \citet[][hereafter HSR98]{hsr98} published a
competing model within the CDM framework that described DLAs a merging
protogalactic clumps.  Their hybrid analysis combined results from a
small set of `zoom-in' numerical simulations with a Press-Schecter
treatment of the dark matter (DM) halo mass function, and a heuristic
assumption for the DM halo dependence of DLA gas cross-section.
HSR98 concluded that this model could reproduce the DLA kinematics
provided that low mass DM halos $(\mvc < 30 \mkms$) [Check; could be 50]
did not host DLAs \citep[see][for a modern update of this
model]{bh09}.

Many other models have since been proposed with a range of success
\citep[e.g.][]{nbf98,mps+01,schaye01a,raz+08,pgp+08}.  Most recently,
\citet[][hereafter H10]{hong+10} proposed that galactic-scale winds
are an essential ingredient to explaining DLA kinematics.  In
reviewing this paper, we noted two treatments of the simulated spectra
that depart from the prescription applied to the DLA observations (as
detailed in PW97).  To our surprise, we discovered that these
mistreatments were first introduced by HSR98.  We expect that H10
simply repeated the HSR98 analysis without realizing the errors.  As 
H10 themselves demonstrate, these have important consequences for the
results derived from the simulations.
With this posting, we wish to highlight these issues in the hopes of
preventing their further propagation.  

\begin{figure}
\begin{center}
\includegraphics[width=3.5in]{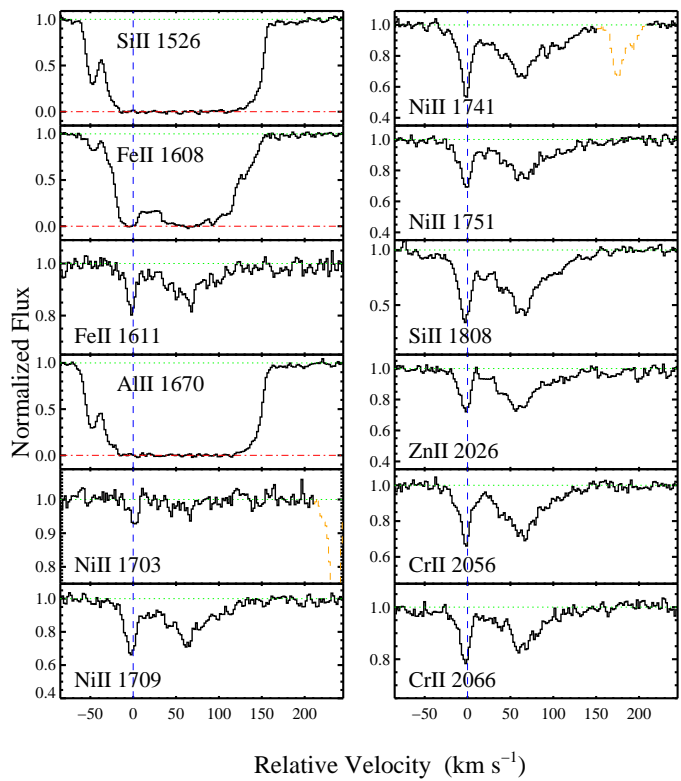}
\end{center}
\caption{
Low-ion transitions for the DLA at $z=1.920$ toward Q2206$-$19
\citep{pw97b}.  Note how closely the various transitions trace one
another, whether or not they arise from the same ion.  Analysis of any
of the lines that satisfy the criterion in Equation~\ref{eqn:depth}
(only \ion{Si}{2}~1808, \ion{Ni}{2}~1741 do) will yield nearly identical
kinematic characteristics.  
At this S/N, one could even include weaker transitions (e.g.\
\ion{Ni}{2}~1751, \ion{Cr}{2}~2056).  We have never precluded
kinematic analysis of a DLA based on this criterion; it only serves to
determine which transition is used.
}
\label{fig:lines}
\end{figure}

\section{The Depth Criterion}

Because the \ion{H}{1} Lyman series of DLAs is highly saturated, we
cannot assess the gas kinematics with neutral hydrogen.  Instead, DLA
kinematics have been assessed using low-ion, metal-line transitions
(e.g.\ \ion{Si}{2}~1808).  To minimize the effects of line-saturation
and statistical noise in the spectra, we imposed a criterion on the
depth of the transition that is analyzed:

\begin{equation}
0.1 \le {\rm min} \ltp \frac{I}{I_c} \rtp \le 0.6
\label{eqn:depth}
\end{equation}
where $I$ is the observed intensity and $I_c$ is the continuum of the
quasar local to the transition.  This criterion precludes lines that
are heavily saturated (i.e.\ $I/I_c < 0.1$) 
also weak profiles to focus the analysis on the bulk of the gas while
maintaing a good S/N ratio (see PW97 for more details).
Conveniently, nature provides enough UV transitions at $\lambda <
2000$\AA\ such that we have always identified at least one low-ion
transition for analysis in every DLA observed, no matter its
\ion{H}{1} column density and metallicity.  
Furthermore, the low-ion transitions very precisely trace one another
\citep[Figure~\ref{fig:lines}][see also]{pro03} such that all lines
satisfying this criterion have nearly identical kinematic
characteristics.  To emphasize, no DLA has
been precluded from kinematic analysis because of this criterion.
Indeed, the original PW97 analysis relied on transitions with a wide
range of $\lambda f$ values (e.g.\ \ion{Fe}{2}~1608,
\ion{Si}{2}~1808).

HSR98 generated metal-line absorption profiles from their numerical
simulations (i) by assuming the \ion{H}{1} gas had a metallicity
[Si/H]~=$-1$ everywhere and (ii) by adopting the atomic data for
the \ion{Si}{2}~1808 transition.  These authors then only considered
profiles ``satisfying the criterion in equation (\ref{eqn:depth})''.
In other words, the \ion{Si}{2}~1808 profiles that violated this
criterion were excised from the analysis.  In fact, this likely
includes nearly all of the profiles that they presented in their
Figures~3-6 as representative of their analysis.  H10 followed the
same steps as HSR98 and thereby excised $30-60\%$ of the profiles from
analysis.
\begin{figure}
\begin{center}
\includegraphics[height=3.5in,angle=90]{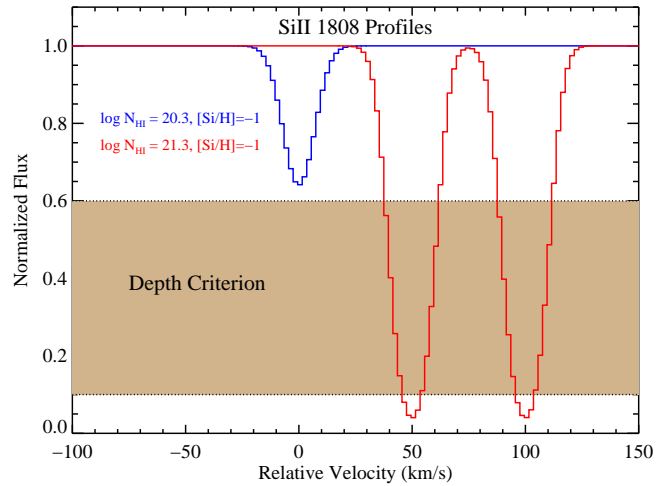}
\end{center}
\caption{
Simulated \ion{Si}{2}~1808 profiles for (blue) a DLA with $\mnhi =
10^{20.3} \cm{-2}$, a metallicity [Si/H]~$= -1$, and a single
component with Doppler parameter $b = 8 \mkms$; and (red) 
LA with total $\mnhi = 10^{21.3} \cm{-2}$, a metallicity [Si/H]~$=
-1$, and two components each with $b = 8 \mkms$.  The shaded region
shows the depth criterion described by Equation~\ref{eqn:depth}.
Neither of these profiles satisfy the criterion:  the low \nhi\ system
is too weak and the high \nhi\ has too saturated of a \ion{Si}{2}~1808
profile.  These profiles would have been excised in the analysis of
HSR98 and H10.  In the observational analysis, however, we would have
identified other low-ion transitions that do satisfy the criterion and
would not have excised the sightline from kinematic analysis.
}
\label{fig:1808}
\end{figure}

It is impossible for us to precisely predict the implications of this
mistreatment, but there are some obvious biases.
Figure~\ref{fig:1808} shows the \ion{Si}{2}~1808 profile for a DLA at
the \ion{H}{1} column density threshold $\mnhi = 10^{20.3} \cm{-2}$
with metallicity [Si/H]~$=-1$, and a single component with Doppler
parameter $b = 8 \mkms$.  We also indicate the depth criterion
expressed in Equation~\ref{eqn:depth}.  In a mistreatment of this
criterion, the sightline would have been excised from kinematic
analysis.  In a proper treatment, one would identify another, strong
transition (e.g.\ \ion{Si}{2}~1526, \ion{Fe}{2}~1608) to perform the
analysis.  In short, the mistreatment biases analysis against
sightlines with low \nhi\ value.  Figure~\ref{fig:1808} also shows   
that sightlines with very large \nhi\ value may produce 
saturated profiles that also violate the depth criterion.  Therefore,
this mistreatment also biases the results against sightlines with very
high \nhi.  Of course, nature provides many more low \nhi\ sightlines
than high \nhi\ sightlines such that the overall bias is against the
former.   The effects kinematically are more difficult to predict.

\section{A \dv\ Threshold}

The other mistreatment of the simulated profiles by HSR98 was:
``For the discussion of velocity widths a minimum threshold of $\mdv >
30 \mkms$ was imposed on both the observed and simulated velocity widths
to avoid incompleteness effects.''.  Observationally, one has never
imposed such a cut.  There is no equivalent to the magnitude limit
in absorption line studies, especially for kinematic
analysis\footnote{It has been oft debated that a limit exists for
  metallicity measurements owing to dust obscuration by the DLA of the
  background quasar.  This would almost certainly bias one against
  sightlines with large velocity width, not small.}.  Every DLA known
that has been observed at even moderate S/N and spectral resolution
shows several low-ion, metal transitions.  Indeed, this has
established the so-called `floor' in metallicity at $\approx 1/1000$
solar for DLAs \citep{pgw+03,pps+10}.  

Imposing a $\mdv < 30 \mkms$ threshold 
actually has a minor effect on the DLA observations because so few
systems exhibit such low \dv\ values \citep[see][for the
latest compilation]{wgp05}.  In contrast, models may produce a
preponderance of systems with small \dv\ values.  
Adopting a \dv\ threshold, therefore, can greatly affect the results
when one compares the model predictions against the observations.
Indeed, the KS probability of the favored wind model in H10 decreases
from 39$\%$ (fine agreement) to 2.8\%\ (nearly ruled out) as they
reduced the threshold from $30 \mkms$ to $20 \mkms$.  It is quite
possible that their model would be ruled out at $>99\%$ c.l.\ if no
threshold were imposed (i.e.\ the proper treatment).  We can only
speculate on the implications for HSR98.  It seems likely that the
primary result from their analysis of the numerical simulations, that
$\mdv \sim 0.6 v_c$, would be greatly diminished (e.g.\ examine their
Figure~10a).  Presumably, this could be tested in the `no wind' models of H10.

\section{Concluding Remarks}

Before concluding, we (sheepishly) apologize for not having identified
these issues previously.   We also fully acknowledge that the many
successes of CDM will not be overturned by simple mistreatments of DLA
kinematic analysis.  Nevertheless, the results appear to remain a
challenge to modern theories of galaxy formation.  Hopefully the
tension that lies between the observations and theoretical prediction
will reveal key aspects of ISM physics and feedback in the early
universe.

\acknowledgments

JXP and AMW are supported by NSF grant (AST-0709235).  JXP
acknowledges the
hospitality of UC San Diego and CASS where he is spending a
sabbatical.


\end{document}